\documentclass{timgad}
\usepackage{amssymb} 
\usepackage[english]{babel}
\usepackage{amsmath}
\usepackage{amsfonts}
\usepackage{amsthm}
\usepackage{hyperref}
\usepackage{float}
\usepackage{enumitem}
\usepackage{bm}
\usepackage[ruled,vlined]{algorithm2e}

\newcommand{\Oh}{\ensuremath{\mathcal{O}}}
\DeclareMathOperator{\td}{td}
\DeclareMathOperator{\wcol}{wcol}

\let\tw\relax
\DeclareMathOperator{\tw}{tw}

\DeclareMathOperator{\SReach}{SReach}
\DeclareMathOperator{\WReach}{WReach}
\DeclareMathOperator{\parent}{\texttt{parent}}
\DeclareMathOperator{\ordering}{\texttt{ordering}}

\usepackage[backend=bibtex,bibencoding=ascii,style=alphabetic,bibstyle=alphabetic,firstinits=true,doi=false,isbn=false,url=false,maxbibnames=99]{biblatex}
\renewbibmacro{in:}{}
\AtEveryBibitem{\clearname{editor}}

\newbibmacro{string+link}[1]{%
	\iffieldundef{ee}
	{\iffieldundef{url}
		{\iffieldundef{doi}
			{\iffieldundef{eprint}
				{#1}
				{\href{http://arxiv.org/abs/\thefield{eprint}}{#1}}}
			{\href{http://dx.doi.org/\thefield{doi}}{#1}}}
		{\href{\thefield{url}}{#1}}}
	{\href{\thefield{ee}}{#1}}} 
\DeclareFieldFormat{title}{\usebibmacro{string+link}{\mkbibemph{#1}}}
\DeclareFieldFormat[article,inproceedings,inbook,incollection,mastersthesis,thesis]{title}{\usebibmacro{string+link}{\mkbibquote{#1}}}

\title{Sallow: a heuristic algorithm for treedepth decompositions}
\author{Marcin Wrochna}
\affil{University of Oxford}
\date{June 2020}

\begin{document}

\maketitle

\begin{abstract}
	\noindent
	\makebox[13em]{Code:}
	\href{https://github.com/marcinwrochna/sallow/releases/tag/v1.0}{github.com/marcinwrochna/sallow/releases/tag/v1.0}\\
	\makebox[13em]{Code DOI:}
	\href{http://doi.org/10.5281/zenodo.3870565}{10.5281/zenodo.3870565}	
	
	\medskip
	We describe a heuristic algorithm for computing treedepth decompositions, submitted for the \href{https://pacechallenge.org/2020}{PACE 2020} challenge. It relies on a variety of greedy algorithms computing elimination orderings, as well as a Divide \& Conquer approach on balanced cuts obtained using a from-scratch reimplementation of the 2016 FlowCutter algorithm by Hamann \& Strasser~\cite{HamannS18}.
\end{abstract}

\section{Orderings and elimination}
We start by recalling a few useful notions and facts (experts will recognize we are essentially describing the well-known statement that $\td(G) = \wcol_\infty(G)$, see e.g.~\cite[Lemma 6.5]{sparsityTreedepth}).

Treedepth has many equivalent definitions.
Small treedepth can be certified as usual by a treedepth decomposition (also known as a Trémaux tree) -- it suffices to specify the $\parent$ of each vertex in the tree.
A corresponding $\ordering$ is any linear ordering of vertices such that parents come before children -- it can be obtained from a $\parent$ vector by any topological sorting algorithm, for example.
In turn, any linear $\ordering$ of vertices can be turned into a treedepth decomposition by an elimination process:
repeatedly remove the last vertex in the ordering and turn its neighbourhood into a clique.
The $\parent$ of the removed vertex is set to the latest vertex in the neighbourhood.

It is easy to check this results in a new valid treedepth decomposition.
Moreover, turning a decomposition into an ordering and back cannot increase the depth.
To see this, observe that by induction, at any point in the elimination process,
the neighbourhood of the removed vertex consists only of its ancestors (in the original decomposition),
because all later vertices were removed,
hence all introduced edges are still in the ancestor-descendant relationship.
This implies in particular that the new parent of each vertex is an ancestor in the original decomposition.

A more static look at the elimination process is through \emph{strongly} and \emph{weakly} reachable vertices.
Fix a vertex $v$.
A vertex $x$ is in the neighbourhood of $v$ at the moment $v$ is eliminated
if and only if $x$ is earlier in the ordering and can be reached, in the original graph, via a path whose internal vertices are later than $v$ in the ordering (= have been eliminated).
We say $x$ is \emph{strongly reachable} from $v$ (in the given graph and ordering).
So this neighbourhood is the set of strongly reachable vertices (in the graph $G$ with ordering $L$), usually denoted $\SReach_{\infty}[G,L,v]$.
Similarly $x$ is \emph{weakly reachable} from $v$ if $x$ is earlier and can be reached via a path whose internal vertices are later than $x$ (instead of ``later than $v$'').
Equivalently, this relation is the transitive closure of strong reachability.
The set of weakly reachable vertices is denoted $\WReach_{\infty}[G,L,v]$ -- it is equal to the set of ancestors of $v$ in the treedepth decomposition obtained by elimination.

To give some context,
the maximum size of $\SReach_{\infty}[G,L,v]$ over vertices $v$ is the \emph{strong $\infty$-colouring number} of $(G,L)$
and its minimum over all orderings $L$ is equal to $\tw(G) + 1$~\cite[Theorem 3.1]{Arnborg85}.
The maximum size of $\WReach_{\infty}[G,L,v]$ over vertices $v$ (hence the depth of the resulting treedepth decomposition) is the \emph{weak $\infty$-colouring number} of $(G,L)$
and its minimum over all orderings $L$ is equal to $\td(G)$.
The $\infty$ here is customary because considering only paths of length at most $k$ leads to similar definitions of strong and weak $k$-colouring numbers, which are important in the theory of sparse graphs, see e.g.~\cite{HeuvelMQRS17}.

A third, more efficient look at the elimination process comes from the observation that 
descendants of $v$, in the resulting treedepth decomposition, are exactly vertices reachable in the subgraph induced by vertices later than or equal to $v$ (in the ordering).
We can thus define a \emph{building process} on an ordering as follows:
we process vertices starting from the last, maintaining connected components of the subgraph induced by vertices processed so far.
This is sufficient to build the same treedepth decomposition, without changing the graph:
we maintain the treedepth decomposition of the subgraph induced by processed vertices
(using a $\parent$ vector)
and represent each component by the root of the corresponding tree (equivalently, the earliest vertex of the component).
When processing a vertex $v$, 
for each neighbour $y$ later than $v$ in the ordering, to update components we only have to
merge $y$'s component with $v$ (initially a singleton).
To update the decomposition,
we find the root of $y$'s component and make it a child of~$v$,
which thus becomes the new root.
Thus $y$'s parent is the latest weakly reachable vertex, as expected.

In other words, the building process consider vertices in the same order as the elimination process.
However, in the elimination process we maintain a graph on unprocessed vertices and when eliminating the next vertex, we replace its neighborhood by a clique, which is somewhat costly.
Instead, in the building process we maintain a graph on more and more processed vertices (hence the name), or rather a structure to represent its connected components.

\section{Greedy algorithms}
The elimination and building processes suggest heuristics for finding treedepth decomposition.
For example, we can start with vertices ordered decreasingly by any notion of centrality (e.g. the degree in the original graph), since we expect higher vertices in optimal decompositions to be more `central'.
Moreover, we can update this `centrality score' of unprocessed vertices on the fly: we maintain a heap of unprocessed vertices with the minimum score at the top, popping and processing a vertex until the heap is empty.
\subsection{By elimination}
In the elimination process, we update the score of a vertex $v$ based on (a linear combination of) 
\begin{itemize}
\item its height (one plus the max height of neighbours eliminated so far; once we decide to eliminate $v$ this becomes the height of the subtree rooted at $v$ in the resulting decomposition);
\item its degree (in the partially eliminated graph; once we decide to eliminate $v$ this becomes $|\SReach_{\infty}[G,L,v]|$ in the resulting ordering);
\item some initial, static score.
\end{itemize} 

Consider a graph with $n$ vertices, $m$ edges, and suppose we stop when unable to obtain a decomposition of depth better than $d$.
In the elimination process, we can then assume that the neighbourhood of every vertex at every step is smaller than $d$.
Simulating it then requires $\Theta(nd^2)$ time in many cases: e.g. if most vertices have a neighbourhood of size $\Theta(d)$ at the time they are processed (as in $K_{d,n}$), ensuring that this neighbourhood becomes a clique takes $\Theta(d^2)$ time.
This bound is also sufficient; to do the simulation we maintain neighbourhood lists of the partially eliminated graph
as \texttt{std::vector}s sorted by vertex name (not by the ordering we're about to compute), so that the union of neighbourhoods can be computed by merge-sort (when turning $v$'s neighbourhood into a clique).
Note however that we cannot compute parents on the fly, since the ordering of unprocessed vertices is not yet decided;
we do this in a second pass, using the faster building approach once the ordering is fixed.
See Algorithm~\ref{alg:elim} below.

The memory requirement is $\Theta(nd)$ in the worst case and this cannot be improved to $\Oh(m)$, because a random $3$-regular graph on $d$ vertices will have, after eliminating half of its vertices, a clique on $\Theta(d)$ vertices and $\Theta(d^2)$ edges (due to its expansion properties).
This means memory usage can also be prohibitive for large graphs with expected treedepth $d$ much larger than the average degree.

For the initial score, the final implementation only uses the height of vertices in the best previously obtained decomposition.
For the parameters $\alpha,\beta,\gamma$ that give the linear combination defining the score,
the best choice for a single run seems to be to choose non-zero $\alpha$, much larger $\beta$, while $\gamma$ can be zero (e.g. $\langle1,9,0\rangle$).
On the other hand, repeating the algorithm with a variety of different parameters often finds significantly better solutions.

\SetKw{KwForIn}{in}
\begin{algorithm}[h]
 \For{$v$ \KwForIn $V(G)$}{
 	$\texttt{g}[v]$ := copy of $N_G(v)$\\
 }
 \texttt{ordering} := $\langle\rangle$\\
 \texttt{height} := $\langle 1,\dots,1\rangle$\\ 
 initialize heap with $v \quad \mapsto \quad \alpha \cdot |\texttt{g}[v]|\ +\ \beta \cdot \texttt{height}[v]\ +\ \gamma \cdot \texttt{score}[v]$\\
 \While{\normalfont heap not empty}{
  $v$ := pop minimum element from top of heap\\
  \texttt{ordering} := $v$, \texttt{ordering}\\
  \For{$x$ \KwForIn{} $\normalfont\texttt{g}[v]$}{
	$\texttt{g}[x]$ := $(\texttt{g}[x] - v) \cup (\texttt{g}[v] - x)$\\
	$\texttt{height}[x]$ := $\max(\texttt{height}[x],\ \texttt{height}[v] + 1)$\\	
	heap.update($x \ \mapsto \  \alpha \cdot |\texttt{g}[x]| + \beta \cdot \texttt{height}[x] + \gamma \cdot \texttt{score}[x]$)\\
  }
  clear $\texttt{g}[v]$\\  
 }
 \texttt{root} := first in \texttt{ordering}\\ 
 \texttt{depth} := $\texttt{height}[\texttt{root}]$\\
 compute \texttt{parent} from \texttt{ordering}\\ 
 \caption{Greedy by elimination on graph $G$ with parameters $\alpha,\beta,\gamma$ and initial \texttt{score}\hspace*{-2em}}
 \label{alg:elim}
\end{algorithm}

\pagebreak[3] 

\subsection{By building}
The $\Theta(nd^2)$ time and $\Theta(nd)$ space bound of the elimination process can be prohibitive for huge graphs of large treedepth.
Instead, the building process can be simulated in $\Oh(\min(m \cdot \alpha(n), n d))$ time and $\Oh(m)$ space by maintaining components with the classic union-find data structure (where $\alpha$ is the inverse Ackerman function~\cite{TarjanL84} and the latter bound follows from the fact that for each vertex, its pointer in the structure only goes up the tree).
We note that this also allows to check the correctness of a treedepth decomposition (by replacing the assignment $\parent[y] := v$ with whatever the original parent was and checking it is a descendant of $v$) in the same running time;
this can be significantly faster than the straightforward $\Oh(md)$ method when $d$ is large.

The details are similar as in Algorithm~\ref{alg:elim}, see Algorithm~\ref{alg:build}.
One change is that $\texttt{g}[v]$ does not represent the neighbourhood of a vertex after elimination;
instead, it represents the graph after contracting processed components.
For a root vertex $r$ of a component $C$ (starting from singleton components), $\texttt{g}[r]$ stores the neighbours of that component.
For a non-root vertex $\texttt{g}[v]$ is cleared (it is important to actually free the memory using \texttt{std::vector::shrink\_to\_fit()}; calling \texttt{clear()} keeps the capacity unchanged).
This guarantees that the total size never increases.
This also means the $\alpha$ part of the score is less meaningful;
using $\alpha \cdot |g[x]|$ below would be the same as $\alpha \cdot |N_G(x)|$ (the original degree, since $x$ is not processed yet).
Instead we use $\alpha \cdot \max(|g[x]|, |g[v]|)$ as a slightly better heuristic.
This does result in noticeably worse results compared to the elimination version.

Moreover, we maintain a union-find structure with pointers $\texttt{ancestor}[v]$.
We decided not to balance unions by size or rank, instead of opting for the more natural choice:
$\texttt{ancestor}[v]$ is always some ancestor in the treedepth decomposition computed so far
($\texttt{ancestor}[v]$ is $\bot$ for unprocessed vertices).
This theoretically spoils the $m \alpha(n)$ running time guarantee, but simplifies the implementation, and we expect the trees to be shallow anyway.

\begin{algorithm}[h]
 \For{$v$ \KwForIn $V(G)$}{
 	$\texttt{g}[v]$ := copy of $N_G(v)$\\
 }
 \texttt{ordering} := $\langle\rangle$\\
 \texttt{parent} := $\langle\bot,\dots,\bot\rangle$\\ 
 \texttt{ancestor} := $\langle\bot,\dots,\bot\rangle$\\  
 \texttt{height} := $\langle 1,\dots,1\rangle$\\ 
 initialize heap with $v \quad \mapsto \quad \alpha \cdot |\texttt{g}[v]|\ +\ \beta \cdot \texttt{height}[v]\ +\ \gamma \cdot \texttt{score}[v]$\\
 \While{\normalfont heap not empty}{
  $v$ := pop minimum element from top of heap\\
  \texttt{ordering} := $v$, \texttt{ordering}\\
  $\texttt{ancestor}[v]$ := $v$\\  
  \For{$y$ \KwForIn{} $N_G(v)$}{
  	\If{$\normalfont\texttt{ancestor[y]} \neq \bot$}{
  		$r$ := find($y$) in the \texttt{ancestor} union-find structure\\
  		\eIf{$r \neq v$}{
  			$\texttt{parent}[r]$ := $v$\\
  			$\texttt{ancestor}[r]$ := $v$ \quad (union)\\
			$\texttt{g}[v]$ := $(\texttt{g}[r] - v) \cup (\texttt{g}[v] - y)$\\
			clear $\texttt{g}[r]$
  		}{
	  		$\texttt{g}[v]$ := $\texttt{g}[v] - y$\\
  		}
		
	}
  }
  \For{$x$ \KwForIn{} $\normalfont\texttt{g}[v]$}{  
	$\texttt{height}[x]$ := $\max(\texttt{height}[x],\ \texttt{height}[v] + 1)$\\	  
 	heap.update($x \ \mapsto \  \alpha \cdot \max(|\texttt{g}[x]|,|\texttt{g}[v]|) + \beta \cdot \texttt{height}[x] + \gamma \cdot \texttt{score}[x]$)\\	
  }
 }
 \caption{Greedy by building on graph $G$ with parameters $\alpha,\beta,\gamma$ and initial \texttt{score}\hspace*{-2em}}
 \label{alg:build}
\end{algorithm}

\subsection{Super-fast version with lookahead}
In Algorithm~\ref{alg:build}  the cost of computing $\texttt{g}[v]$ is still significant (though much lower compared to the elimination version).
A super-fast version can be obtained by removing $\texttt{g}[v]$; however the $\texttt{height}$ of unprocessed vertices cannot be maintained exactly then.
In that case the heap is useless and we can simply do the building process with a fixed ordering by initial score.

However, we can significantly improve this super-fast version with a simple lookahead.
Instead of processing the last unprocessed vertex, we check the last $\ell$ unprocessed vertices, compute what their height would be at this point, and choose the minimum height.
For $\ell=2$ this is almost as fast as a DFS;
for $\ell=64$ this is still faster than other versions and results in significantly better depth than DFS, often giving a reasonable ballpark estimate.
Nevertheless we essentially always use the full version of Algorithm~\ref{alg:build} as well, unless we know that the super-fast estimates are good enough (e.g. in recursive runs where other branches are already deeper).

\subsection{By building with lookahead}
A similar idea can be used to get the best of the elimination and building versions.
The problem with the building version is that we do not have access to the degree of a vertex after eliminations, for evaluating the heuristic score.
A work-around is to do this evaluation exactly (by computing unions of neighbourhoods) for a few vertices close to the top of the heap.
In fact, a re-evaluation can only increase the score, pushing a vertex down,
so it suffices to re-evaluate and update the top vertex of the heap some constant $\ell$ number of times (completely forgetting the computed unions of neighbourhoods afterwards).
We can stop as soon as the top vertex stays at the top after re-evaluation, so even high constants $\ell$ turn out to be quite affordable.
For $\ell=1024$, this results in an algorithm that seems just as good as greedy by elimination, yet avoids the heavy memory usage in huge graphs of large treedepth.

All in all, on the public instances of the PACE 2020 challenge, the total score (sum of $\frac{\text{best known depth}}{\text{algorithm's depth}}$ over all tests) and total running time was roughly (with $\langle\alpha,\beta,\gamma\rangle=\langle 1,9,0\rangle$):
\begin{itemize}[itemsep=0pt,parsep=-1ex]
\item 37\% in 1 minute for the ``super-fast greedy'' version with $\ell=64$,\\
\item 63\% in 8 minutes for the basic ``greedy by building'' version (Algorithm~\ref{alg:build}),\\
\item 72\% in 15 minutes for the ``building with lookahead $\ell=1024$'' version,\\
\item 86\% in 3000 minutes for the ``building with lookahead $\ell=1024$'' version running 30 minutes on each instance with various $\langle\alpha,\beta,\gamma\rangle$,\\
\item 96\% in 3000 minutes for the final algorithm with Divide \& Conquer.
\end{itemize}
(single-thread on a i7-8550U CPU with \texttt{-O3 -march=ivybridge -flto}).

\section{Divide \& Conquer}
The other main component of the submitted algorithm is a simple divide \& conquer idea: find a possibly small balanced cut, remove it, recurse into connected components, and output a treedepth decomposition with the cut arranged in a line above the recursively obtained decompositions.

To find balanced cuts we use the FlowCutter algorithm submitted for the PACE 2016 challenge by Ben Strasser.
It is a crucial part here as well, but the idea and details are already very well described in a paper by Hamman and Strasser~\cite{HamannS18} (see also some further details in~\cite{Strasser17}).
In this submission it is only perhaps noteworthy that the algorithm has been reimplemented from scratch,
in an attempt to optimize it (it is after all the main bottleneck on most large instances).
The only semantic difference however is that the new implementation works directly on vertex cuts and vertex flows.
Unfortunately, this adds some technical complexity, since for example augmenting paths of vertex flows can revisit a vertex twice.
An actual experimental comparison remains to be done, but this seems to matter less than how the cuts are actually used.

Due to time constraints of the author, the final algorithm uses the greedy and FlowCutter components in suboptimal ways, in a rather fragile and unprincipled patchwork.
It starts with greedy algorithms and then takes the first cut of at least a given balance yielded by FlowCutter and recurses.
Afterwards FlowCutter is ran again, without using the fact that in one run it can output several cuts of increasing balance and size.
Recursive results are also never used again, instead the algorithm just runs in a loop with more and more costly recursions.

We attempted modifying FlowCutter to judge balance not based on sizes of the two sides, but on estimates of their treedepth (using callbacks).
It appears this did not bring significant improvements to the final result, so this modification is turned off in the submitted version.

To speed things up a final feature is that of cutoffs.
A bad cutoff $d$ tells the algorithm to abandon any attempt that won't lead to a decomposition of depth strictly smaller than $d$.
For example, we use the best know decomposition's depth as a bad cutoff most of the time.
A good cutoff $d$ tells the algorithm to return as soon as it can output a decomposition of depth at most~$d$.
For example, we use the maximum depth in sibling branches computed so far.
The estimates obtained at early phases of the algorithm are also used to cut recursion attempts that are unlikely to yield better decompositions.
We also use a lower bound based on the degeneracy of a graph and the longest path found in a DFS to quickly finish in some easy cases.

\subsection*{Acknowledgements}
The author is very grateful to the \href{https://pacechallenge.org/2020}{PACE 2020} organizers at the University of Warsaw, organizers of past editions, and the OPTIL.io team at the Poznań University of Technology for making this challenge possible.
This project has received funding from the European Research Council (ERC) under the European Union’s Horizon 2020 research and innovation programme (grant agreement No 714532, PI: Stanislav \v{Z}ivn\'{y}).
\begin{picture}(0,0)
\put(392,10)
{\hbox{\includegraphics[width=40px]{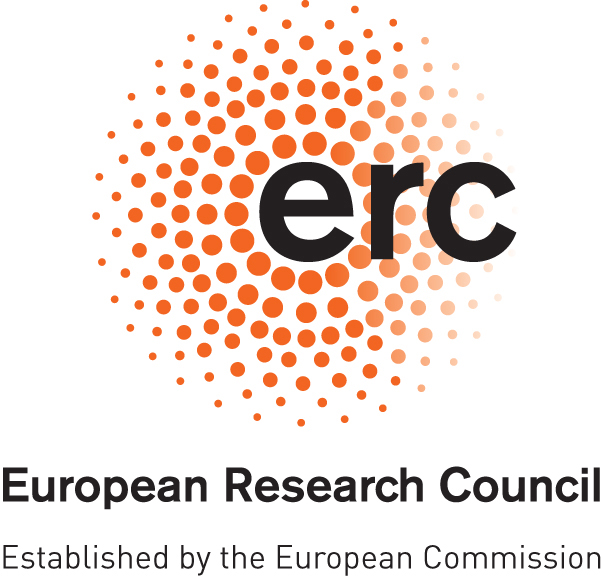}}}
\put(382,-50)
{\hbox{\includegraphics[width=60px]{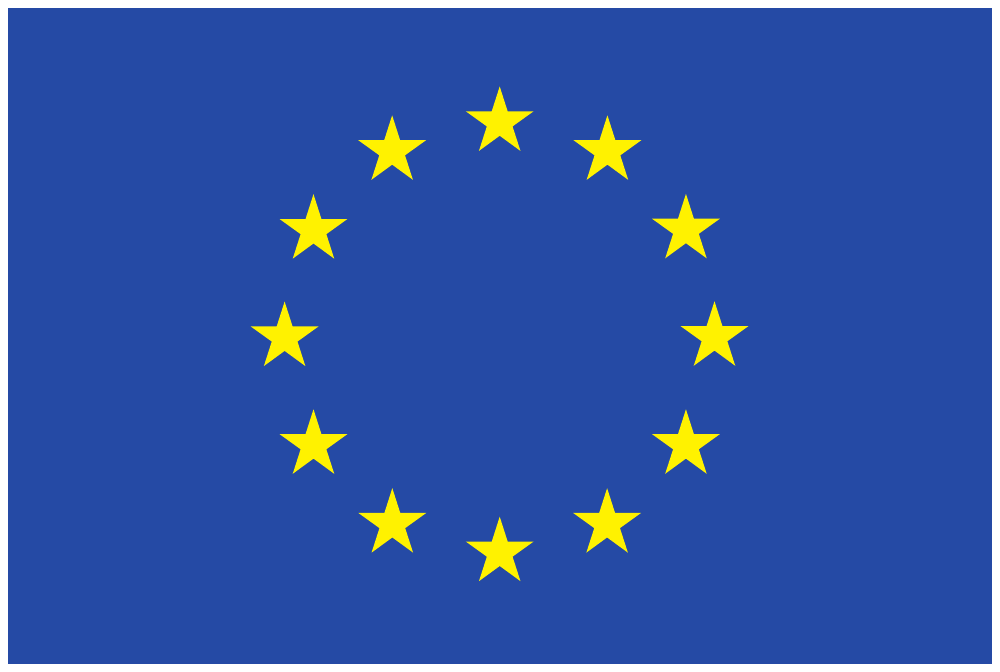}}}
\end{picture}


\printbibliography

\end{document}